\documentclass[fleqn,usenatbib]{mnras}

\usepackage{graphicx}	
\usepackage{amsmath}	
\usepackage{amssymb}	

\title{The impact of ionizing radiation on the formation of a supermassive star in the early Universe}

\author[S. Chon et al.]{
Sunmyon Chon,$^{1}$\thanks{E-mail: sunmyon.chon@utap.phys.s.u-tokyo.ac.jp}
and Muhammad A. Latif\,$^{2,3}$
\\
$^{1}$Department of Physics, School of Science, University of Tokyo, Bunkyo, Tokyo 113-0033, Japan\\
$^{2}$Sorbonne Universit\`es, UPMC Univ Paris 06 et CNRS, UMR 7095, Institut d'Astrophysique de Paris, 98 bis bd Arago, 75014 Paris, France \\
$^{3}$ Department of Physics,  COMSATS Institute of Information Technology, Park Road, Islamabad 44000, Pakistan
}

\date{Accepted XXX. Received YYY; in original form ZZZ}

\pubyear{2016}

\begin{document}
\label{firstpage}
\pagerange{\pageref{firstpage}--\pageref{lastpage}}
\maketitle

\begin{abstract}
A  massive primordial halo near an intensely star-forming galaxy may collapse into a supermassive star (SMS) and leave a  massive black hole seed of about $10^5~M_{\odot}$.  To investigate the impact of ionizing radiation on the formation of an SMS  from a nearby galaxy, we perform three-dimensional radiation hydrodynamical simulations by selecting a pair of  massive dark matter halos forming at $z >10$. We find that rich structures such as clumps and filaments around the source galaxy shield the cloud from ionizing radiation. In fact, in some cases, cloud collapse is accelerated under ionizing radiation.
This fact suggests that the ionization of the cloud's surroundings helps its collapse. Only strong radiation at the early stage of structure formation can halt the cloud collapse, but this is much stronger than observationally allowed value.  We also explored the effect of ionizing radiation on a sample of 68 halos by employing an analytical model and found that increase in the mean density of the gas between the SMS forming cloud and the source galaxy protects the gas cloud from ionizing  radiation as they approach each other. Thus, we conclude that ionizing radiation does not prevent the formation of  an SMS in most of the cases.
\end{abstract}

\section{Introduction}
The observations of high redshift quasars reveal the existence of supermassive black holes (SMBHs) at  $z > 6$ \citep{Mortlock+2011,Wu+2015}.  SMBHs have typical masses of a few billion solar and the Eddington limited accretion requires that their seeds should be more massive than the observed stellar mass BHs  in the local Universe \citep{Belczynski+2010,Casares+2014,GW150914}.

One possible pathway to form a high-$z$ SMBH is  the remnant BH from Population III (Pop III) stars, which have the masses in the range of $10$--$1000~M_\odot$ \citep{Heger+2003, Susa+2014, Hirano+2014, Hirano+2015, Hosokawa+2016}. In this model, the observed mass can only be attained when the Eddington accretion is maintained over about one billion years. However, the radiation feedback initially from the Pop~III and later from the BH itself impedes mass accretion and stops its growth \citep{Yoshida2006, Milos+2009, PR2011}.

The direct collapse (DC) scenario is another promising way to form SMBHs, potentially first forming a supermassive star (SMS) of $10^5$--$10^6~M_\odot$ that later collapses into an equal mass BH \citep{Shibata+2002}
and grows to the observed masses by rapid mass accretion \citep{Volonteri2010, DiMatteo+2012}.  The SMS is expected to form out of a pristine cloud with almost no H$_2$ \citep{BL03,Inayoshi+2014,Sakurai+2016},
which is the main coolant in normal primordial  gas clouds. The photons  with energy between $11.2$ and $13.6~\mathrm{eV}$, so-called Lyman--Werner (LW) radiation, directly dissociate molecular hydrogen \citep{StecherWilliams1967} while the low energy (above 0.76 eV) photons photodetach $\rm H^-$, which impedes the major $\rm H_2$ formation channel in the early Universe.  A massive primordial gas cloud in the presence of a strong LW flux cools by atomic  hydrogen lines and collapses almost isothermally with $T\sim 8000~\mathrm{K}$ \citep{Omukai2001}. To enable the atomic line cooling, the gas cloud should  be hosted by a primordial DM halo with $T_\text{vir} > 8000~$K dubbed as `DC halo'.

The conditions for the formation of  an SMS (free from metal and H$_2$ cooling) have been studied by various authors finding that in the presence of a strong far-ultraviolet (FUV) radiation, the formation of molecular hydrogen remains suppressed \citep{Regan+2009b,Regan+2009a,Latif+2013,  Becerra+2015,Choi+2015, Latif+2016}, while other mechanisms to prevent H$_2$ cooling are also proposed \citep{Inayoshi+2012,Tanaka+2013,Fernandez+2014}. The critical value of the required FUV flux is found to be much larger than the background and such a flux can be achieved in the close vicinity of a star-forming galaxy \citep{Dijkstra+2008,Dijkstra+2014, Visbal+2014b, Latif+2015}.  \cite{Chon+2016}  performed  cosmological simulations coupled with a semi-analytical model and found that the dynamical effects, e.g. tidal force from the FUV source galaxy or ram pressure stripping, are also important for the formation of an SMS. It is because the source halos are located in the close vicinity of the DC halo.  They showed that an SMS can form by avoiding the possible tidal disruption from the FUV source galaxy.

The FUV source galaxy also emits  ionizing radiation that can affect the formation of an SMS in several ways.  The ionizing radiation can promote the formation of H$_2$ by increasing the degree of ionization and may elevate the strength of a critical flux  required to dissociate H$_2$ formation \citep{Johnson+2014,Inayoshi+2015}.  They can also photoevaporate the DC halo by injecting thermal energy and may halt the formation of an SMS. 
This makes the DC halo more prone to the photoheating by the source galaxy. These effects need to be quantified by 3D cosmological simulations  coupled with radiative transfer. 

\cite{Regan+2016a,Regan+2016b} performed radiation hydrodynamical simulations to study the impact of ionizing radiation on the DC scenario. They found that the SMS-forming cloud irradiated by ionizing photons gets photoevaporated before its collapse. Only clouds located  far from the star forming galaxy can collapse while the  H$_2$ cooling  is not completely suppressed in such a cloud. 
They arbitrarily placed the radiation source at the distance of a few kpc from the DC halo and employed an approximate ISM model to take into account the attenuation of ionizing photons.
However, they  may overestimate the impact of ionizing radiation as the ionizing source did not correspond to the density peak of underlying matter. In the real universe,  galaxies are expected to form at  the center of  DM halos  embedded in the cosmic web and surrounded by  rich structures such as filaments and walls. 
Moreover, they assumed a constant luminosity throughout the simulation. Doing so, they overestimated the impact of radiation at earlier stages of galaxy evolution.

In this study, we investigate the impact of ionizing radiation on the DC halo from a nearby star forming galaxy under a realistic configuration by taking a halo pair from cosmological simulations of \cite{Chon+2016}. Hydrodynamical simulations coupled with radiative transfer are performed by placing the radiation source at the center of the star forming halo and  the source luminosity is computed from the star formation history by employing the semi-analytical model.  We compute time dependent source luminosity by taking into account the halo growth.
Our results suggest  that the DC halo remains unionized for a realistic source luminosity owing to the absorption of ionizing radiation in the dense filaments around the source halo. We also estimated the impact of ionizing radiation on 68 DC halos taken from  \cite{Chon+2016} and found that  most of them remain unaffected.

This paper is organized as follows. In section 2, we describe numerical methods and simulations setup. In section 3, we present the results from radiation hydrodynamical simulations. The effect of the ionizing radiation on the DC halo in broader context is discussed in Section 4. Finally,  we confer our conclusions in section 5.

\section{Methodology}
We use $N$-body + smoothed particle hydrodynamics (SPH) code, Gadget-3 \citep{Springel2005} to perform cosmological hydrodynamical simulations and our simulation setup is based on \cite{Chon+2016}. Here, we briefly summarize  the main features and discuss newly added physics. We use the following cosmological parameters provided by  the $Planck$ data  \citep{PlanckXVI2014},  $\Omega_\text{m} = 0.308$, $\Omega_\text{b} = 0.0483$, $\Omega_\Lambda = 0.692$, $h = 0.677$, and $\sigma_8 = 0.8288$. 

\subsection{$N$-body simulations}
We first perform $N$-body simulations starting from cosmological initial conditions  generated with the MUSIC \citep{HahnAble2013} at $z = 99$  which employs second-order Lagrange perturbation theory. The box size in our base simulation is $20~h^{-1}~$Mpc and $128^3$ DM particles are used to compute dark matter dynamics.  We select 10 most massive halos at $z \sim 9$ using the friends-of-friends (FOF) algorithm and re-run simulations with higher resolution in the identified halos. Each zoom-in region has a volume of $1.2\;h^{-1}\;\mathrm{Mpc}^3$, and the particle mass resolution of  $\sim 1.8\times 10^2\;h^{-1}\;M_\odot$. Mini-halos of $\sim 10^5\;\mathrm{M_\odot}$ are resolved by more than 100 particles in our simulation.

The physical properties of halos are defined as follows \citep{Barkana+2001}:
\begin{align}
T_\text{vir} &= 1.98 \times 10^4 \left ( \frac{\mu}{0.6} \right ) 
\left [ \frac{\Omega_\text{m}}{\Omega_\text{m}(z)} 
\frac{\Delta_\text{c}}{18\pi^2} \right ]^{1/3} \nonumber \\
&\;\;\;\; \left ( \frac{M}{10^8\;h^{-1}\;M_\odot} \right )^{2/3} \left ( \frac{1+z}{10} \right ) \;\;\;\; \mathrm{K}, \label{eq_Tvir} \\
R_\text{vir} &= 0.784 \left [ \frac{\Omega_\text{m}}{\Omega_\text{m}(z)} 
\frac{\Delta_\text{c}}{18\pi^2} \right ]^{-1/3}  \nonumber \\ 
&\;\;\;\; \left ( \frac{M}{10^8\;h^{-1}\;M_\odot} \right ) ^{1/3}  \left ( \frac{1+z}{10} \right )^{-1} h^{-1}\;\mathrm{kpc}, \\
V_\text{c} &= 23.4 \left [ \frac{\Omega_\text{m}}{\Omega_\text{m}(z)} \frac{\Delta_\text{c}}{18\pi^2} \right ]^{1/6} \nonumber \\
&\;\;\;\; \left ( \frac{M}{10^8\;h^{-1}\;M_\odot} \right ) ^{1/3}  \left ( \frac{1+z}{10} \right )^{1/2} \mathrm{km\;s^{-1}},
\end{align} 
where $\mu$ is the mean molecular weight, $\Omega_\text{m} (z)$  the matter density, and $\Delta_\text{c}$  the overdensity of the halo.

\subsection{Semi-analytic Model for Galaxy formation} \label{sec_SAM}
 In our model,  recipes for the star formation and metal enrichment in the halo are computed by mainly following \cite{Agarwal+2012} and are discussed in detail in  \cite{Chon+2016}. The halo is labelled as  `metal enriched' once its virial temperature  exceeds $2000$~K, at which Pop III star formation is expected to occur inside the halo \citep{Tegmark+1997}. Under LW radiation, we modify the critical mass for the Pop III star formation ($M_\text{crit}$)  as \citep{Machacek+2001};
\begin{equation}
M_\text{crit} = \psi \left ( 1.25 \times 10^5 + 2.8 \times 10^6 \;J_\text{21}^{0.47} \right ) M_\odot,
\end{equation}
where $\psi \simeq 4$ is the non-dimensional parameter introduced by \cite{ON2008} and $J_\text{21}$ is the local LW intensity in the unit of $10^{-21}~\mathrm{erg~s^{-1}~Hz^{-1}~cm^{-2}~sr^{-1}}$.

We allow continuous Pop II star formation in the halos with $ M > 10^7~M_\odot$ and compute star formation history in each halo from the semi-analytical model described below. 
In our model,  we  include gas cooling, star formation, and feedback from SNe and divide  baryons into three components, hot gas, cold gas, and stars. We assume that the hot gas cools over dynamical time ($t_\text{dyn}$) and turns into the cold gas that later forms stars over the star formation time-scale ($t_\text{SF}$). Here, $t_\text{dyn} \equiv R_\text{vir}/V_\text{c}$ and $t_\text{SF} \equiv 0.1t_\text{dyn}/\alpha$, where $\alpha$ is the star formation efficiency.
Based on the star formation history, we calculate the energy spectrum of each halo using the population synthesis code, STARBURST99 \citep{Leitherer+1999}.

We select DC halos based on the following criteria;
\begin{enumerate}
\setlength{\itemsep}{.5mm}
\item the halo is pristine;
\item the halo is irradiated by a strong LW radiation with $J_{21} > J^\text{crit}_{21}$;
\item the halo has $T_\text{vir} > 8000~$K, massive enough for  atomic cooling to operate.
\end{enumerate}
Here, we take $J^\text{crit}_{21} = 100$ assuming radiation spectra of Pop~II stars to be the blackbody with $T_\text{eff} = 10^4~$K \citep{Shang+2010}. We do not consider LW radiation originating from Pop~III stars because they have shorter lifetimes and are also less abundant than Pop~II stars.

\subsection{Hydrodynamical simulation} \label{sec::setup_hydro}
We use Gadget-3 for hydrodynamical simulations and couple it with our chemical model.  The rate equations of the following 14 species ($\mathrm{e}^{-}$, $\mathrm{H}$, $\mathrm{H}^+$, $\mathrm{He}$, $\mathrm{He}^{+}$, $\mathrm{He}^{2+}$, $\mathrm{H_2}$, $\mathrm{H_2}^{+}$, $\mathrm{H}^{-}$, $\mathrm{D}$, $\mathrm{D}^{+}$, $\mathrm{HD}$, $\mathrm{HD}^{+}$, and $\mathrm{D}^{-}$) are solved with an implicit scheme. The reaction network is taken from \cite{Yoshida+2003, Yoshida+2006}.
We also incorporate chemical reactions induced by stellar radiation. We divide the radiation into two components, low energy ($h\nu < 13.6~\mathrm{eV}$) and ionizing ($h\nu > 13.6~\mathrm{eV}$) radiation.
The former component photodissociates H$_2$ and HD molecules as well as photodetaches $\text{H}^-$ while the latter photoionizes H.

We create a sink particle at the local minimum of the gravitational potential  once the gas particle density exceeds $10^3\;\mathrm{cm}^{-3}$ \citep{Hubber+2013}. 
We adopt the sink radius of $1\;\mathrm{comoving\;kpc}$, which is comparable to the size of a star cluster \citep{Wise+2012}.
The sink particle grows in mass by accreting gas particles within the sink radius. The accreted gas particle is removed from  simulations and  its mass is added to the sink particle. We allow the merger of sink particles if the sink particle enters inside the radius of the other sink particle.

\footnotetext{The virial temperature is calculated from the DM only calculations because the impact of the baryons on the DM distribution is very small.}

\subsection{Radiation Transfer}
We divide the stellar spectrum into two components, low energy  and ionizing radiations. The luminosity in each energy range is assigned to the sink particle after its creation at the galaxy center. Here, we describe how we compute the luminosities and solve the radiation transfer.

\subsubsection{Low energy radiation}
We assume the spectrum of the radiation ($ h\nu < 13.6~$eV) to be blackbody with $T_\text{eff}=10^4~$K. Time dependent LW luminosity ($L_\text{LW}$) is computed from the semi-analytical model assuming optically thin limit ($\propto L_\text{LW}/r^2$, where $r$ is the distance from the source galaxy). In our simulations, we further assume that the LW intensity is spatially uniform inside the DC halo.

\subsubsection{Ionizing radiation}
We consider  photoionization  of hydrogen and take into account the photoelectric heating due to ionizing radiation.
The spectrum of ionizing radiation ($h\nu > 13.6$~eV) is assumed to be the blackbody spectrum with $T_\text{eff}=10^5\;\mathrm{K}$. 
We assign luminosity to the sink particle after its creation by considering four different models named as 1e41, 1e42, SFE0.005, and SFE0.1. For  the  cases of 1e41 and 1e42, the luminosity is fixed at $10^{41}$ and $10^{42 }~\mathrm{erg~s^{-1}}$, respectively.  While in models SFE0.005 and SFE0.1, the luminosity is given by the star formation rate ($\dot{M}_*$) as \citep{Iliev+2005b, Visbal+2016}: 
\begin{equation} \label{eq_luminosity}
L_\text{UV} = 3.3 \times 10^{42} \left ( \frac{\dot{M}_* }{1\;M_\odot~\mathrm{yr}^{-1}} \right ) \mathrm{erg~s^{-1}}.
\end{equation}
Here, $\dot{M}_*$ is  computed from the semi-analytical model discussed in section \ref{sec_SAM} and the star formation efficiency ($\alpha$) is  0.1 and 0.005 for  SFE0.1 and SFE0.005, respectively. For the source luminosity calculation, we assume the escape fraction of ionizing photons to be unity (inside a sink particle), while the escape fraction from the entire galaxy is computed self-consistently in our simulation as we employ radiative transfer.

To perform radiative transfer, we implement ray-tracing scheme proposed by \cite{Susa2006} in Gadget-3. In this scheme,  local optical depth for the `upstream particle'  is computed that is located within the smoothing length of the target particle and is closest to the ray from the radiation source. The total optical depth from the source is obtained by summing the local optical depths. Finally, the ionization rate is given by the photon conserving method \citep{Abel+1999,Kessel-Deynet+2000}.
\\ \\
Our main purpose in this study is to investigate the impact of ionizing radiation on collapse of the DC halo and we note that the ionizing radiation spectrum is inconsistent with the LW spectrum ($T_\text{eff} = 10^4$~K).  Our halo sample is taken from our previous study \citep{Chon+2016} where DC candidate halos are selected assuming  $T_\text{eff}=10^4\;\mathrm{K}$. We therefore keep the same spectrum for LW radiation. On the other hand, for $T_\text{eff}=10^4\;\mathrm{K}$ the radiation spectrum is too soft and  the number of ionizing photons are expected to be much smaller. Thus, we adopt the $T_\text{eff}=10^5\;\mathrm{K}$ blackbody spectrum for radiation with $h\nu > 13.6$~eV. This choice allows us to  investigate the role of ionizing photons during the DC.  In Section~\ref{sec::discussion_spectrum}, we will discuss how the spectrum of the radiation source affects our results.

\begin{figure} 
	\centering
		\includegraphics[width=9cm]{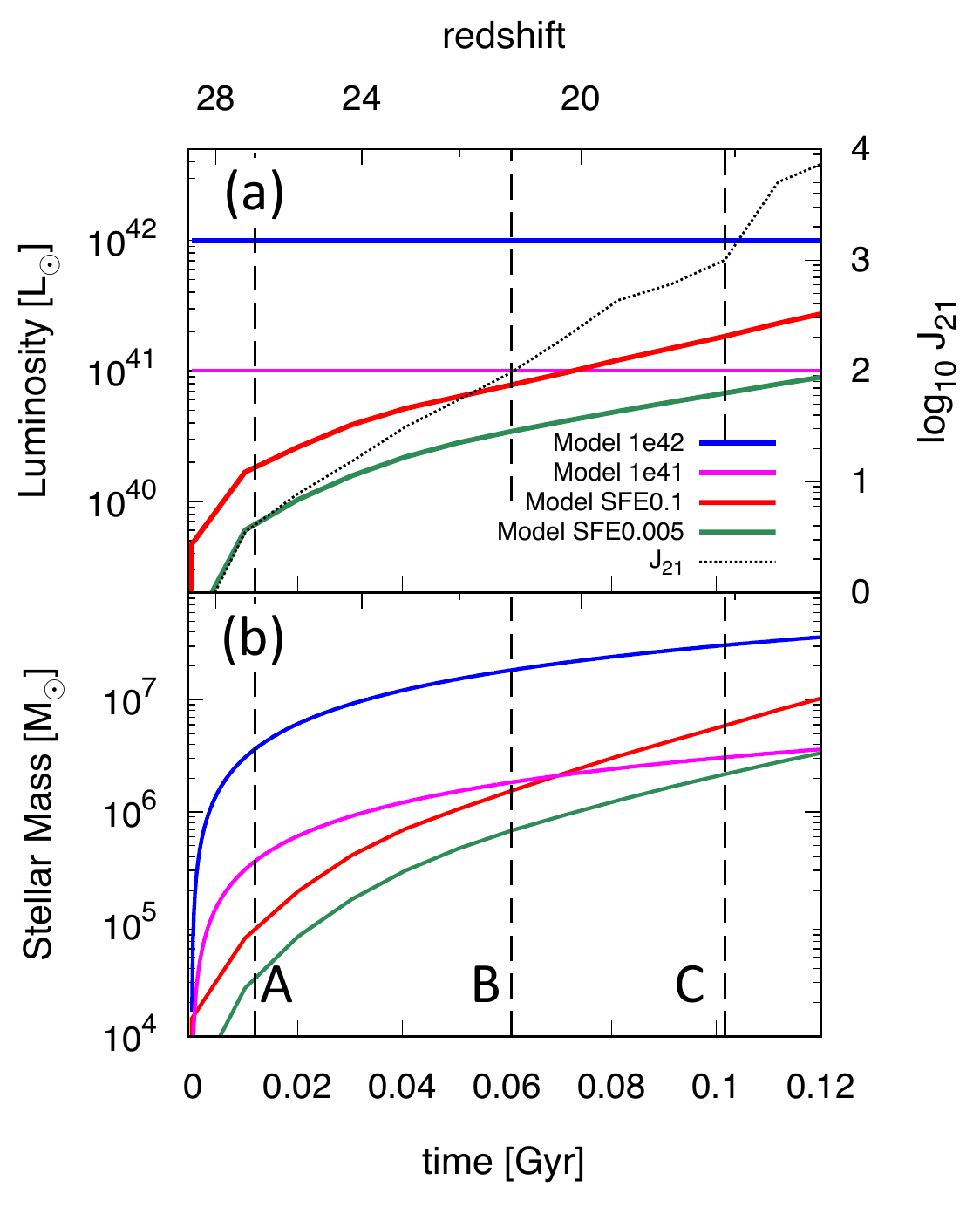}
		\caption{Time evolution of source halo luminosity (panel a) and stellar mass (panel b) for models 1e41 (magenta), 1e42 (blue), SFE0.1 (red), and SFE0.005 (green). Time origin is set to be the moment when the source halo starts to emit ionizing photons. The vertical dashed lines  A, B, and C represent the reference points used in the later discussion. The stellar masses for SFE0.1 and SFE0.005 are calculated from the semi-analytical model while for models 1e41 and 1e42, the stellar masses are given by equation \eqref{eq_luminosity}. The dotted line in panel (a) represents $J_\text{21}$, which is assumed to be the same for all models.
		}
		\label{fig_lum_sfh}
\end{figure}

\begin{figure}
	\centering
		\includegraphics[width=8cm]{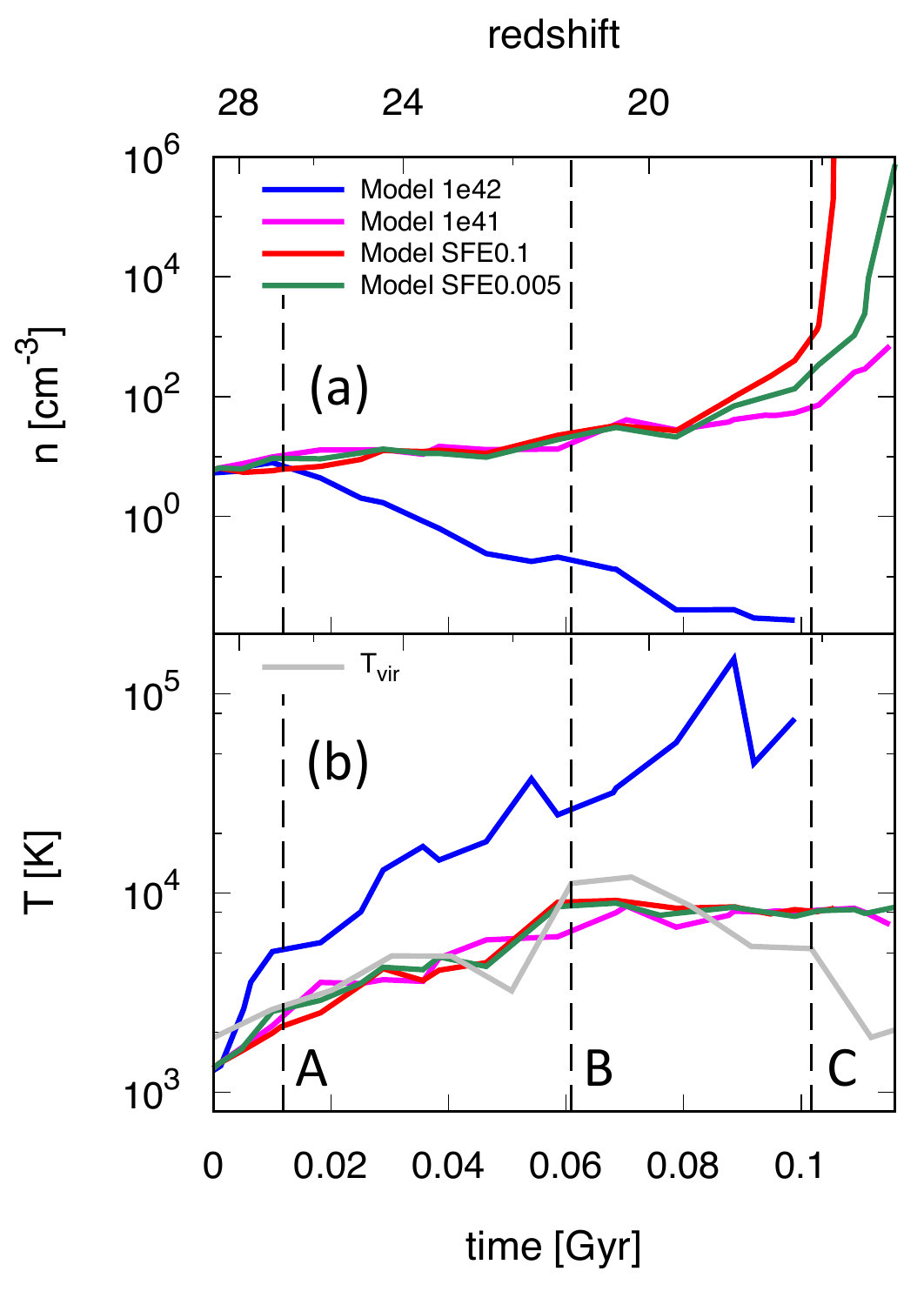}
		\caption{
		Density (panel a)  and temperature evolution (panel b) of the DC halo center for models 
		1e41 (magenta),
 		1e42 (blue),  SFE0.1 (red), and SFE0.005 (green). Time origin is the moment when the galaxy starts to emit ionizing photons. The grey line shows the virial temperature of the host halo of the DC halo.
		 The virial temperature decreases after $t > 0.08~$Gyr because the outer envelop of the halo is stripped by the tidal field originating from the source galaxy. \protect \footnotemark
		}
		\label{fig_density_evolution}
\end{figure}

\section{Results} \label{sec_result} 
We have performed 3D cosmological simulations coupled with radiative transfer to investigate the impact of ionizing photons on the collapse of DC halos for models 1e41, 1e42, SFE0.005, and SFE0.1.  Our main findings are presented here.

 The time evolution of the UV luminosity, LW intensity ($J_{21}$), and the stellar mass in the source galaxy are shown in FIg~\ref{fig_lum_sfh} for all four models. The time origin corresponds to the moment at which the halo mass reaches $10^7~h^{-1}~M_\odot$ and starts to emit ionizing radiation.  
Here, we calculate the stellar mass in models 1e41 and 1e42 by integrating equation \eqref{eq_luminosity}, which is different from that given by the semi-analytical model. 
Both UV luminosity and stellar mass  gradually increase over time due to the growth of halo mass. If we compare models SFE0.1 and SFE0.005, the star formation rate efficiency is  20 times higher in SFE0.1 while the stellar mass  is only 2-5 times  larger compared to the SFE0.005. This  comes from the fact that  the star formation is mainly regulated by cooling of hot gas in Model SFE0.1 and thus the star formation rate saturates at some point.
 
We note that the model 1e42 assumes an extremely large UV luminosity for a given halo mass. The corresponding stellar mass is $10^7~M_\odot$ that is roughly equal to the halo mass at $z \sim 25$.  Nevertheless, we study this case to demonstrate that the strong ionizing radiation at the early stage of the halo formation photoevaporates the halo and  prevents its collapse. We study the model 1e41 for the sake of a comparison with the  work of \cite{Regan+2016a}. The stellar mass for our model 1e41 is $10^6~M_\odot$, an order of magnitude higher than in \cite{Regan+2016a}. The discrepancy mainly comes from the assumed star formation history. The ionizing luminosity is mainly provided by the massive stars that have  lifetimes of $1$--$10~$Myr, much shorter than the duration of our simulation (0.1 Gyr) and the luminosity starts to decrease after massive stars die. Therefore, to maintain a constant luminosity of  $10^{41}~\mathrm{erg~s^{-1}}$ massive stars continuously form in our simulations  and consequently stellar mass is higher in our case. While on the other hand, \cite{Regan+2016a} assume bursty mode of star formation, which is only valid for the stellar lifetime longer than the simulated time.

\begin{figure*}
	\centering
		\includegraphics[width=18.3cm]{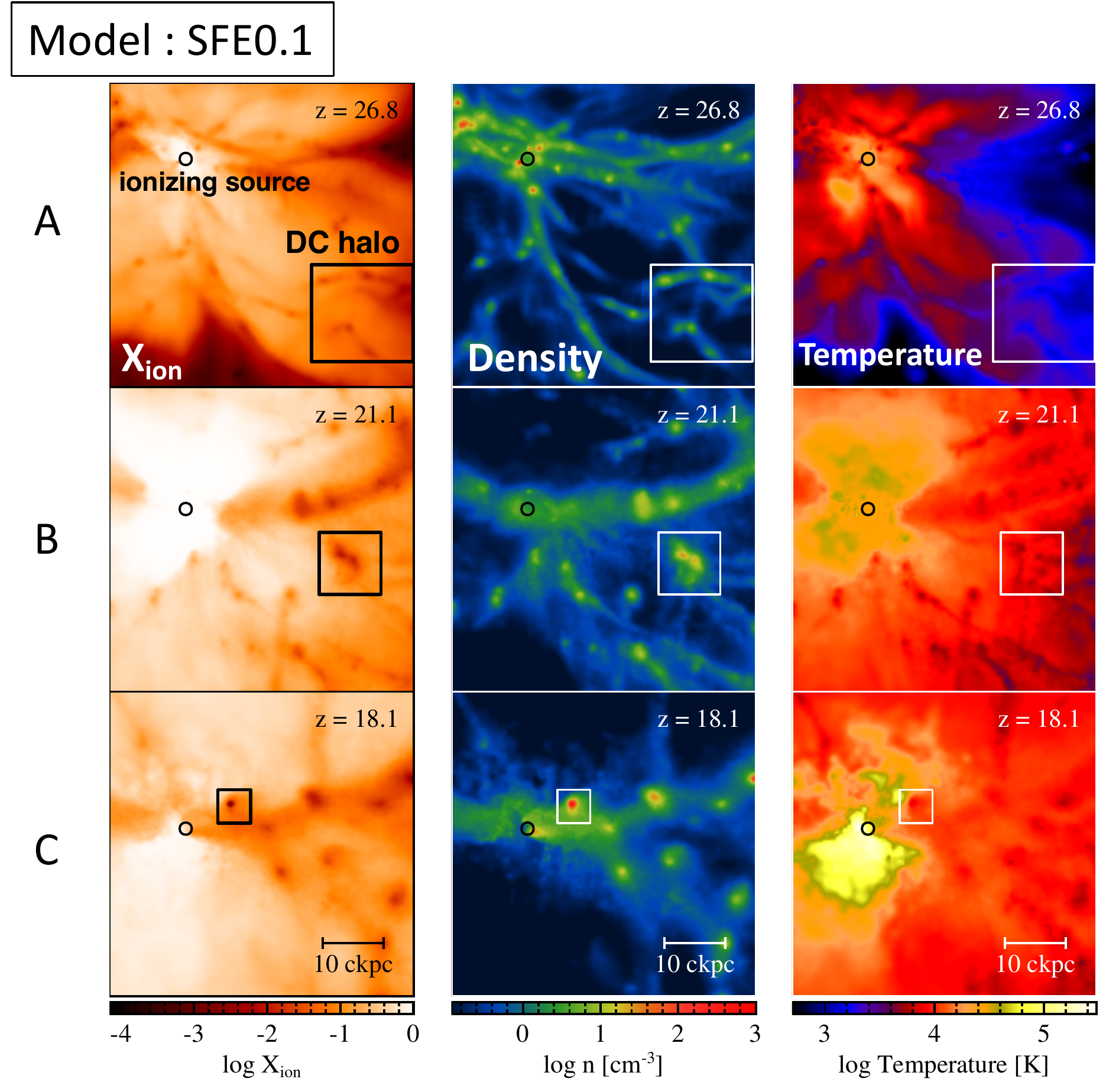}
		\caption{
		We show gas properties for model SFE0.1, $X_\text{ion}$, $n$, and temperature at the left, middle, and right columns, respectively.  Each row (A, B, and C) corresponds to the snapshot at the reference point indicated in Fig. \ref{fig_lum_sfh}. The black filled circle in each panel represents the ionizing source galaxy and the square box shows the focused DC halo or progenitors of the DC halo. The HII region expands in the anisotropic manner reflecting the large scale structure in the Universe. The DC halo is surrounded by the dense filament, therefore remains neutral and collapses into the dense core.
		}
		\label{fig_modelsfe01}
\end{figure*}

\begin{figure}
	\centering
		\includegraphics[width=8cm]{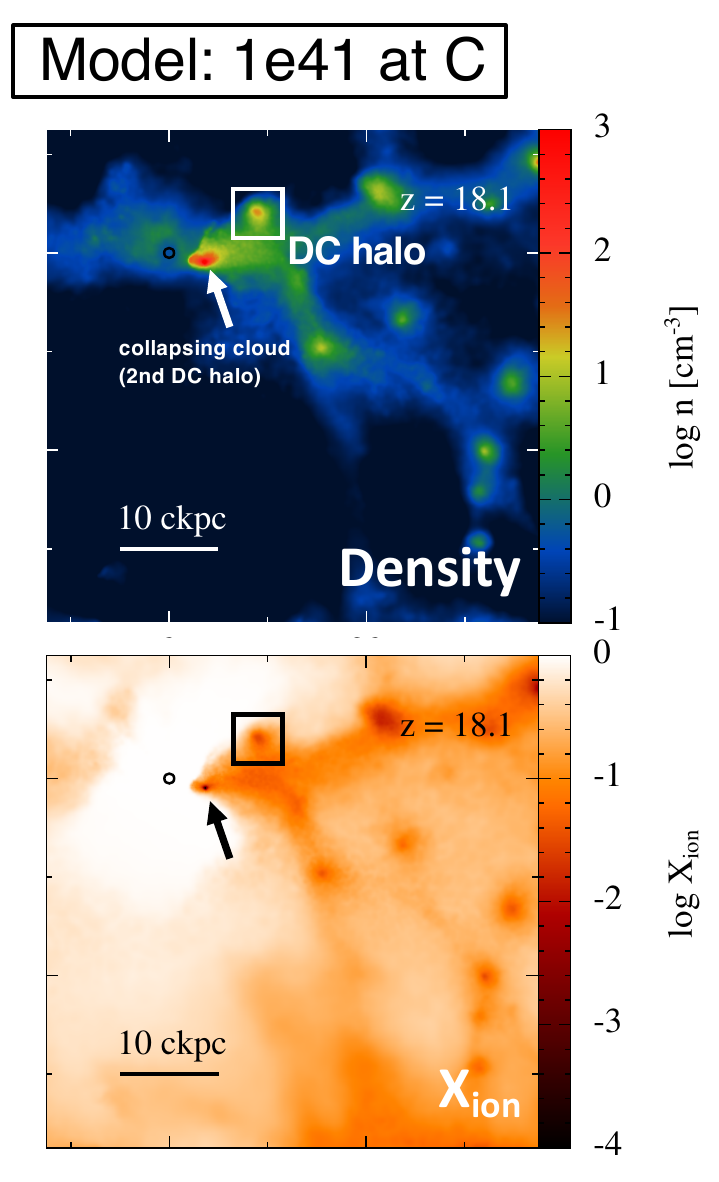}
		\caption{
		Density (top) and $X_\text{ion}$ (bottom) maps for model 1e41 at the reference point C in Fig.~\ref{fig_lum_sfh}. The black circle and square show the position of the light source and the DC cloud, respectively. Thick arrows show the collapsing cloud near the DC halo (second DC halo).
		}
		\label{fig_model1e41}
\end{figure}

\subsection{Cloud collapse} \label{sec_collapse}
To investigate the cloud collapse, we  computed time evolution of the cloud density and temperature as  shown in Fig.~\ref{fig_density_evolution}. The density monotonically increases with time and reaches $10^8~\mathrm{cm}^{-3}$ at $t \sim 0.1~\mathrm{Gyr}$ in models SFE0.1 and SFE0.005 while for  the model 1e42 the density starts to decrease. 
The temperature evolution for models SFE0.1 and SFE0.005 shows that  the gas temperature stays close to the halo virial temperature that indicates that UV heating is inefficient at the halo center.  However, for model 1e42, the DC halo gets heated by ionizing radiation and  the temperature becomes larger than the virial temperature.  Thus, gas no longer remains bounded by the gravitational potential of the host halo and the density starts to decrease.

In model 1e41,  the evolution of density and temperature  during the first  $t\lesssim0.08~$Gyr is similar to  the SFE0.1 and SFE0.005 cases. However, the DC cloud could not collapse to high densities due to its photoevaporation  and finally gets merged with the source halo. However,  interestingly another DC halo forming in the surroundings of the source galaxy is able to collapse  for model 1e41 and is discussed  in the next section.
\begin{figure}
	\centering
		\includegraphics[width=7cm]{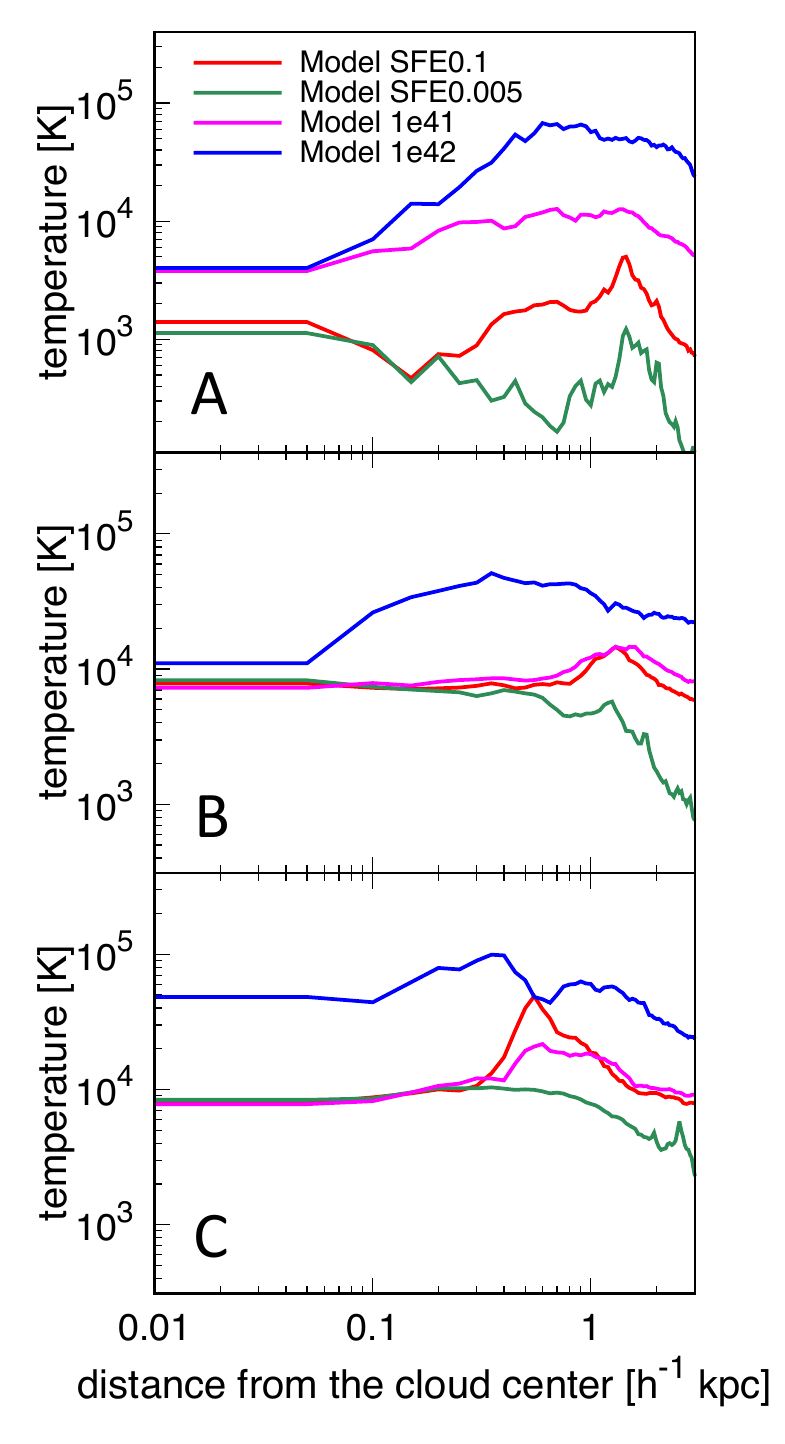}
		\caption{
		Radial temperature profiles around the DC halo for models 1e42 (blue), 1e41 (magenta), SFE0.005 (green), and SFE0.1 (red). Each panel corresponds to the profile at the reference points A, B, and C shown in Fig. \ref{fig_lum_sfh}.
		}
		\label{fig_temperature_profile}
\end{figure}

\begin{figure}
	\centering
	\includegraphics[width=7.3cm]{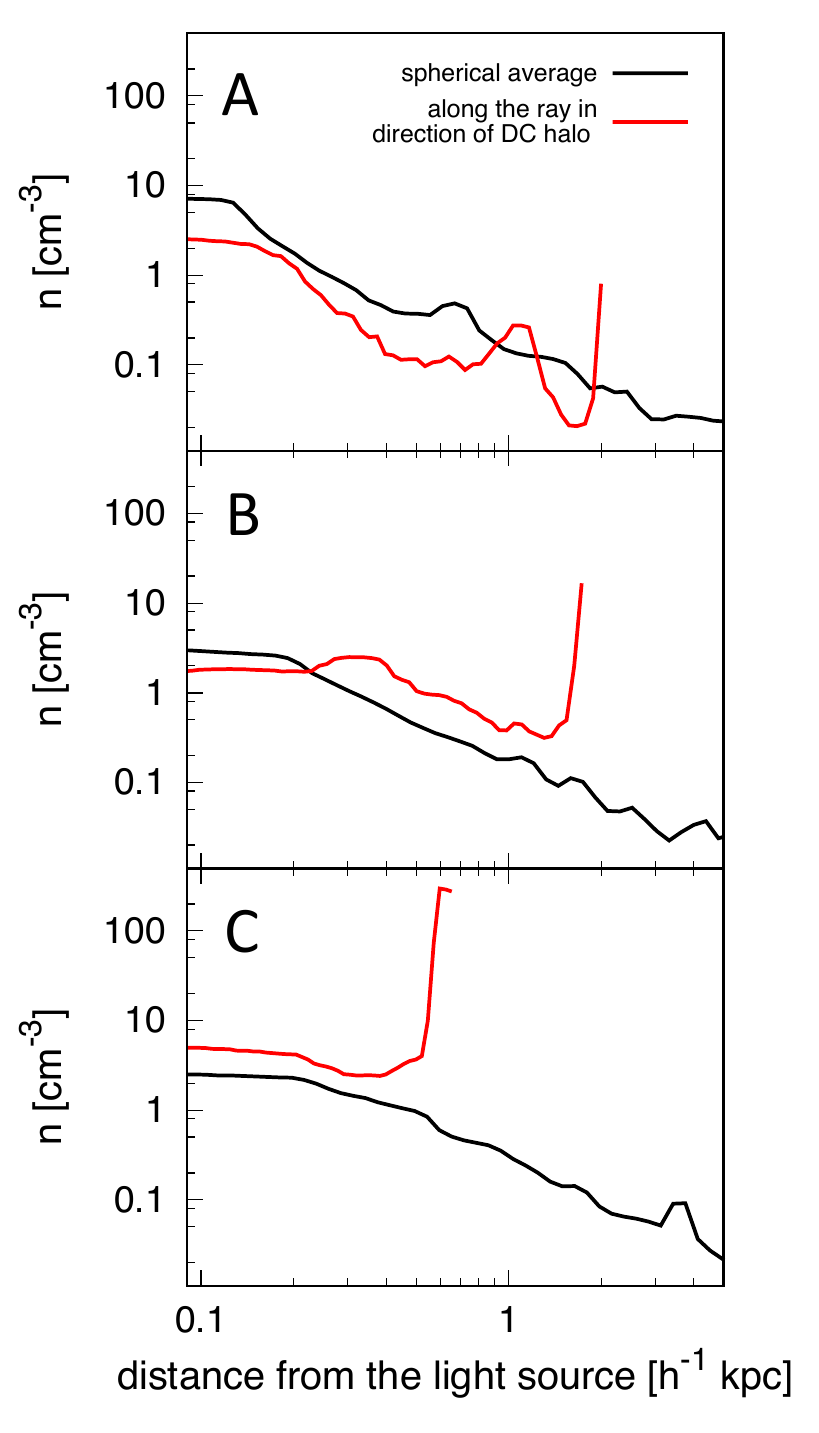}
	\caption{
	Radial density profiles around the source halo for model SFE0.1. The black lines show the spherical averaged profiles while the red lines show the density profiles along the ray that is directed towards the DC cloud. Each panel corresponds to the profile at the reference points A, B, and C  shown in Fig. \ref{fig_lum_sfh}.
	}
	\label{fig_density_profiles}
\end{figure}

The simulations are performed until the central density reaches $10^8~\mathrm{cm}^{-3}$. The cloud collapses into a dense core for models SFE0.1 and SFE0.005 and the density evolves faster in the former case. 
For  the  case of SFE0.1,  stronger UV luminosity  heats  the cloud surroundings to $10^5~$K that may help the cloud collapse.
The  model SFE0.1 assumes strongest UV luminosity compatible with the observation \citep[][and reference therein]{Agarwal+2012} and our results suggest that even in such a case, the DC halo can collapse without being photoevaporated. 

\subsection{Visual Inspection}
The ionization degree ($X_\text{ion}$=$n_\text{HII}/(n_\text{HI} + n_\text{HII})$), the number density ($n$) and the temperature distribution around the source galaxy and DC halo  are shown in Fig.~\ref{fig_modelsfe01} for model SFE0.1.  It  is found that the HII region expands in an anisotropic manner around the source halo  due to the presence of filaments and voids around the source halo. The expansion of the ionization front gets stalled in the filament while it continues to expand in the void region as the source halo is embedded in the growing dense filament (the second column of Fig. \ref{fig_modelsfe01}).

As  DC halo approaches the source galaxy, the dense filament  shields it from ionizing radiation and reduces their impact. The gas temperature within the DC cloud remains around $\sim 10^4$~K, which is much smaller than that of surrounding gas ($\text{several} \times10^4~\mathrm{K}$). The latter even accelerates the cloud collapse (also see Sections \ref{sec_collapse} and \ref{sec_T_profile}) and the  DC halo collapses into the dense core (Fig.~\ref{fig_modelsfe01}C). 
 
In model SFE0.005, the surroundings of the source halo remain almost neutral and the cloud evolves in the same way as without ionizing radiation where the cloud temperature is regulated at $\sim 8000$~K by Ly-$\alpha$ cooling.
In model 1e42, the HII region quickly expands and encompasses the DC halo just after the source galaxy starts to emit ionizing photons. The ambient gas around the source is heated up to $\sim 10^5$~K and such high temperature smoothes out density fluctuations around the source galaxy. Therefore,  the DC halo could not  collapse.

For model 1e41, the density and $X_\text{ion}$ maps at point C are shown in Figure~\ref{fig_model1e41}.  The density at the DC cloud center is  $\sim 10$--$10^2~\mathrm{cm^{-3}}$,  about two orders of magnitude smaller than in the model SFE0.1 at the same redshift. Similarly, the degree of ionization in the surroundings of the source is also higher for model 1e41.  Because of the strong ionizing radiation at early stage, the surroundings of the DC halo is heated up to $\gtrsim 10^4~$K and it gradually loses mass by photoevaporation (see also Fig.~\ref{fig_temperature_profile}) as its progenitors are directly exposed to  ionizing photons. Moreover, the size of HII region around source is larger compared to the SFE0.1. This explains why the DC cloud could not collapse and finally gets merged with the source galaxy. 

Surprisingly, another DC halo (hereafter we call it as second DC halo and the collapsed cloud in model SFE0.1 as first DC halo) forming in the vicinity of source galaxy is able to collapse up to densities of $10^8~\mathrm{cm^{-3}}$ for model 1e41 (highlighted by the arrow in Fig.~\ref{fig_model1e41}).  The second DC halo has different formation history and its progenitors are formed inside the dense filament that kept them shielded from ionizing radiation. The second DC halo is collapsed at the edge of HII region and its collapse is likely accelerated by the thermal compression as the collapse of the second DC halo is not observed in other models.  It is the promising candidate site for SMS formation. This suggests that some halos forming in the vicinity of a strong radiation source with luminosity of $10^{41}~\mathrm{erg~s^{-1}}$ may survive from ionizing radiation and lead to the formation of an SMS.

To further elucidate the difference between two halos for model 1e41, we computed the ionizing flux at various epochs and found that it is almost similar for both halos. We found that the first DC halo  could not collapse because four out of its six progenitors are completely photoevaporated just after epoch A.  On the other hand progenitors of the second DC halo are formed in the dense filament and remain shielded from ionizing radiation. Our results suggest  that the merger history of the halo plays an important role in their collapse.

\subsection{Temperature and  density profiles around the DC halo} \label{sec_T_profile}
The temperature radial profiles around the DC halo  at the reference points A, B, and C  are depicted in Fig.~\ref{fig_temperature_profile}. In model SFE0.005, the DC halo and its ambient are never heated to  $ >10^4$~K which indicates that there is no effect of ionizing photons from the source galaxy. In model SFE0.1,  the temperature profile within the central part ($R < 0.1~\mathrm{kpc}$) of the halo is  similar to model SFE0.005. While in the outer region ($R \sim 1~$kpc), the temperature is  higher than $10^4~$K which may act as a compressive force due to the pressure gradient and help the cloud collapse (Fig.~\ref{fig_density_evolution}a). 

For model 1e42, gas is heated up  to  $ \sim 6000~$K  immediately after the galaxy starts to emit ionizing photons and the gas temperature becomes higher than the virial temperature of the halo. So gas is no longer bounded and starts to escape from the halo.  The gas temperature continues to increase and  exceeds $10^4~$K  at $R\sim 0.1~$kpc. Consequently, the cloud gets evaporated and can no longer collapse.

For  model 1e41, the temperature inside the DC halo is instantaneously increased up to a few times $10^3~$K (Fig.~\ref{fig_temperature_profile} A) and lies between the 1e42 and SFE0.1 cases. As the filament grows around the source galaxy, the ionizing flux at the DC halo decreases.  Consequently, the temperature approaches the model SFE0.1 (Figs.~\ref{fig_temperature_profile}B and C). At epoch C, the rise of temperature between $0.5$ and $1$ kpc corresponds to  the larger luminosity in SFE0.1 case.

We show a spherically averaged density profile and the density along the ray directed towards the DC halo from the source galaxy (we define this as the `ray density') in Fig.~\ref{fig_density_profiles} for SFE0.1 case.  At point A, the ray density between two halos is smaller than the averaged density around the source.  However,  the ray density increases with time and  reaches $\sim 2$--$3~\mathrm{cm}^{-3}$ at  point C while  the averaged density decreases with time. This  explains the decrease in ionization fraction (recombination rate is proportional to the square of the cloud density) as the DC halo approaches the source galaxy.  The HII region only expands in the  void region and filaments remain neutral because of  the higher gas density. Thus once the DC halo has been surrounded by the filament, it can avoid the effect of ionizing photons and safely collapse into a dense core. 

\section{Discussion}
\subsection{Size of HII region} \label{sec_stromgren}
We estimate the size of an HII region ($R_\text{st}$) around the light source from the balance between ionization and recombination of hydrogen:
\begin{align} \label{eq_rst}
R_\text{st} &= \left ( \frac{3 L_\text{UV}}{4\pi\bar{n}^2\alpha_\text{B} E_\text{UV}} \right )^{1/3} \nonumber \\
&= 1.14~\mathrm{kpc} \left ( \frac{L_\text{UV}}{10^{42}~\mathrm{erg~s^{-1}}} \right )^{1/3}
\left ( \frac{\bar{n}}{1~\mathrm{cm}^{-3}} \right )^{-2/3},
\end{align}
where $L_\text{UV}$ is the UV luminosity of the source, $E_\text{UV}$ is the mean energy of ionizing photons, $\bar{n}$ is the mean number density within the HII region, and $\alpha_\text{B}$ is the case-B recombination coefficient of H. 
We here assume $E_\text{UV} = 13.6~$eV independent of the ionizing spectra\footnotemark
\footnotetext{For a realistic ionizing spectra, the mean ionizing energy $E_\text{UV}$ becomes slightly larger than $13.6~$eV (14.6, 16.0, and 35.8~eV for $T_\text{eff}=10^4$, $2\times10^4$, and $10^5~$K blackbody spectra, respectively).}
and estimate the $J_{21}$ as a function of $L_\text{UV}$ as follows:
\begin{align}
J_{21} = \frac{L_\text{LW}}{4\pi^2r^2\Delta \nu} = \frac{\beta L_\text{UV}}{4\pi^2r^2\Delta \nu},
\end{align}
where $L_\text{LW}$ is the luminosity in the LW band, $\Delta \nu$ is the frequency width of the LW band, $\beta$ is the ratio of LW to UV luminosity, and $r$ is the distance from the source. 
$\beta \gtrsim 1$ is typical for  star-forming galaxies (see also Table \ref{tab_spectra}).
For now, we neglect the shielding effect of LW radiation and discuss its implication at the end of the section.  We define the distance at which a halo receives the critical LW intensity $J_{21,\text{crit}}$ as
\begin{align}\label{eq_rJ21}
R_\text{J21} &= \left (   \frac{\beta L_\text{UV}}{4\pi^2\Delta \nu J_{21,\text{crit}}}  \right ) ^ {1/2} \nonumber \\
&= 6.77 ~\mathrm{kpc} ~
\beta^{1/2} \left ( \frac{L_\text{UV}}{10^{42}~\mathrm{erg~s^{-1}}} \right )^{1/2}
\left ( \frac{100}{J_{21,\text{crit}}} \right )^{1/2} .
\end{align}

Equations \eqref{eq_rst} and \eqref{eq_rJ21} show that  $R_\text{st}$ scales as $\propto L^{1/3}$ while $R_\text{J21}$ varies as $\propto L^{1/2}$ under the fixed spectrum. Comparing these equations, we get the critical luminosity $L_\text{c}$, above (below) which $R_\text{J21}$ becomes larger (smaller) than $R_\text{st}$,
\begin{equation}
L_\text{c} = 2.3 \times 10^{37}~\mathrm{erg~s^{-1}} \beta^{-3} \left ( \frac{J_{21,\text{crit}}}{100} \right )^3
\left ( \frac{\bar{n}}{1~\mathrm
{cm}^{-3}} \right )^{-4}. 
\end{equation}
This equation indicates that the critical luminosity strongly depends on $\bar{n}$ and the radiation spectrum.  For the  parameter choice in our simulations ($\beta \sim 1$--$10$ and $J_{21,\text{crit}} = 100$), $L_\text{c}$ remains smaller than the luminosity of the central galaxy $\sim 10^{41}~\mathrm{erg~s^{-1}}$. This is one of the reasons that the DC halo can collapse without being photoevaporated by the ionizing radiation.

To evaluate the shielding effect of LW radiation, we estimate the column density of H$_2$ ($N_{\text{H}_2}$). \cite{Draine+1996} note that  for $N_{\text{H}_2}$  $ >10^{14}~\mathrm{cm}^{-2}$, LW radiation are attenuated for a static medium with no velocity gradients. $N_{\text{H}_2}$ at distance $R_\text{J21}$ can be computed in the following way:
\begin{align} \label{eq_nh2}
N_{\text{H}_2} &=  2.1\times 10^{13} ~\mathrm{cm^{-2}} ~\beta^{1/2} 
\left ( \frac{L_\text{UV}}{10^{42}~\mathrm{erg~s^{-1}}} \right )^{1/2}\nonumber \\ 
&\left ( \frac{100}{J_{21,\text{crit}}} \right )^{1/2} 
\left ( \frac{f_\mathrm{H_2}}{10^{-9}} \right )
\left ( \frac{\bar{n}}{1~\mathrm{cm^{-3}}} \right ) ,
\end{align}
where $f_{\text{H}_2}$ is the $\text{H}_2$ fraction within the HII region. The fiducial value  of $f_{\text{H}_2}=10^{-9}$ is taken from \cite{Omukai2001} and is consistent with our simulation results. Equation \eqref{eq_nh2} shows that the shielding can be neglected unless the spectrum is extremely soft ($\beta \gg 10$).  In the high density region with $n \gg 1~\mathrm{cm}^{-3}$, we may overestimate the LW intensity. Since we are mainly interested in the evaporation of the cloud that has a typical density of $1~\mathrm{cm}^{-3}$, we therefore only consider the parameter regime with $n \lesssim 1~\mathrm{cm}^{-3}$.

The above discussion is only true for the static medium. For a significant velocity gradient of the infalling cloud, larger than the sound speed of the source galaxy,  the Doppler shift of the radiation reduces the effect of shielding \citep{WolcottGreen+2011}. In our case the width of the LW line ($\Delta \nu$) is equivalent to the sound speed ($c_\text{s}$), $\Delta \nu = (c_\text{s}/c) \nu$,  we therefore expect that the shielding effect can be almost neglected.

\begin{figure}
	\centering
	\includegraphics[width=8.6cm]{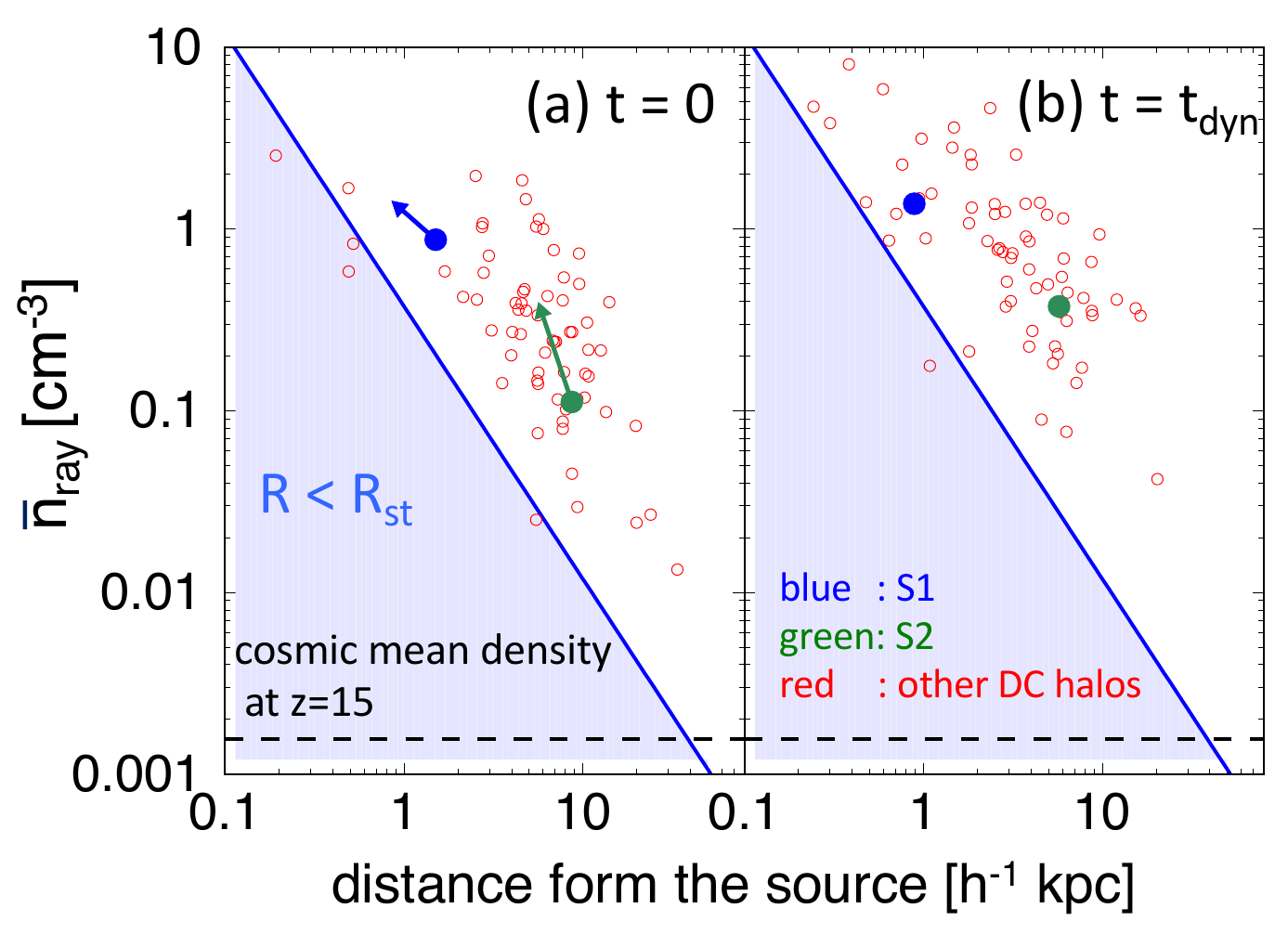}
	\caption{
	 Mean number density along the ray in direction to the DC halo from the source halo ($\bar{n}_\text{ray}$)  is plotted against the distance from the source halo. Panels (a) and (b) show the snapshots at the different epoch, $t = 0$ and $t_\text{dyn}$, where 	time origin is the moment when the halos satisfy the DC criteria. The circles represent all the 68 DC halos found in the N-body simulation and filled circles correspond to the DC halos confirmed to collapse by hydrodynamical simulation. The blue filled circle corresponds to the halo whose evolution is followed by the radiative hydrodynamical simulation in the present study (Fig. \ref{fig_modelsfe01}). The end points of arrows in panel (a) indicate values at $t=t_\text{dyn}$. The dashed lines represent the cosmic mean density at $z=15$ while shaded regions represent $R < R_\text{st}$ for $L_\text{UV} = 10^{41}~\mathrm{erg~s^{-1}}$.
	}
	\label{fig_nmean}
\end{figure}

\begin{figure}
	\centering
	\includegraphics[width=7.3cm]{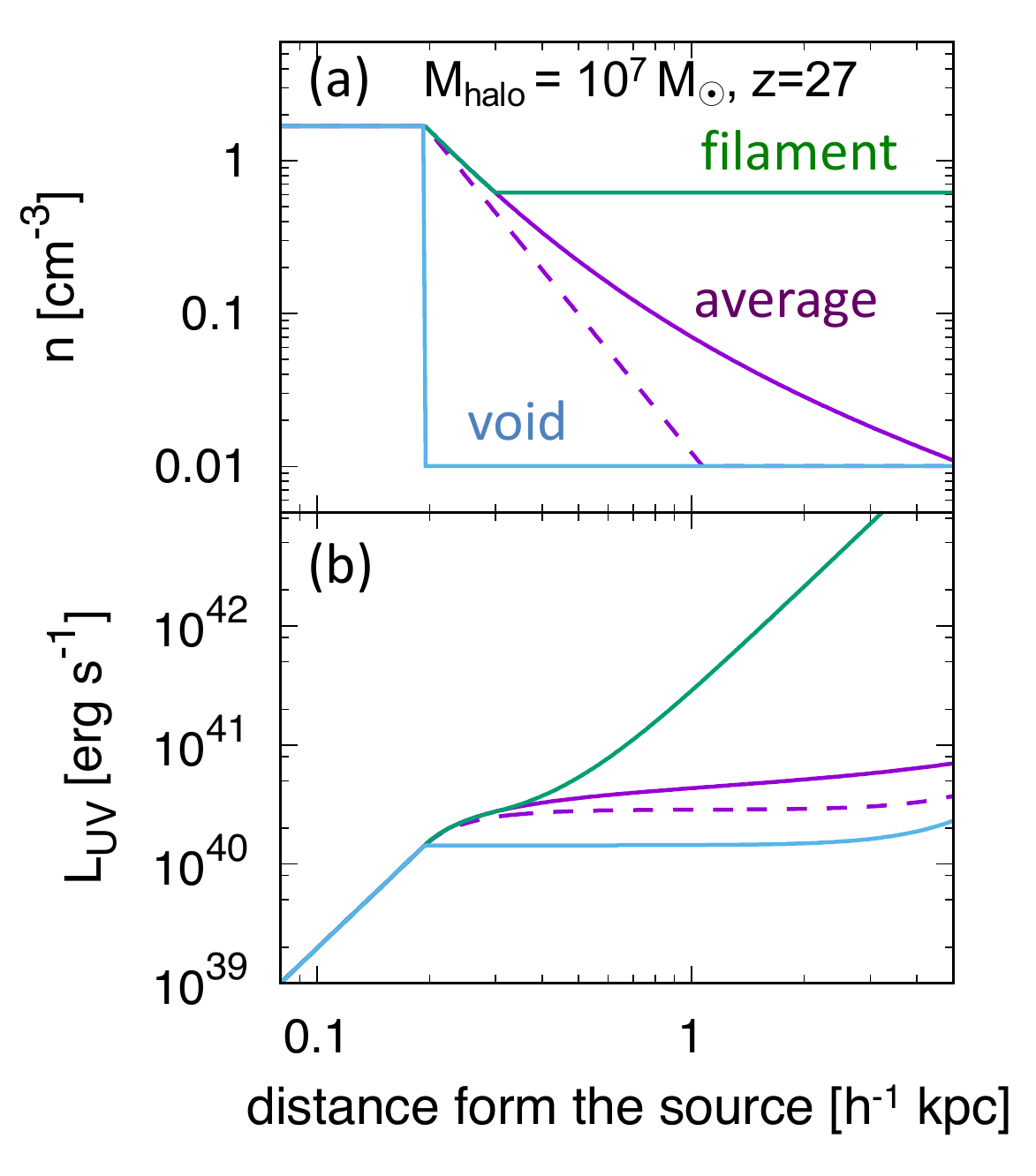}
	\caption{
	(a) Radial density structure around the halo, whose mass is $10^7~M_\odot$ at $z=27$. The purple lines show the mean density profile around the halo, while the green and blue lines show the density structure around the halo  at the filamentary and void regions, respectively. The NFW  and  spherical collapse profiles  are shown by dashed and solid lines, respectively. (b) UV luminosity ($L_\text{UV}$) radial profiles required to ionize hydrogen for the density profiles in the panel a. 
	}
	\label{fig_R_LUV}
\end{figure}

\subsection{Ionization of the DC candidate halos}
We focus on the DC halos selected from $N$-body simulations and discuss the impact of ionizing photons on their evolution. The analysis presented here is based on the discussion in section \ref{sec_stromgren} and in total, we analyze 68 DC halos.  These halos do not always host an SMS  as  \cite{Chon+2016} found from hydrodynamical simulations that 40 out of 42 DC halos could not collapse due to the tidal field originating from a nearby star-forming galaxy. 
We emphasize that they share the similar environment as the two collapsed DC halos.

Fig.~\ref{fig_nmean} shows the mean density and the separation between the DC halo and the light source halo for our sample halos. The mean density $\bar{n}_\text{ray}$ is calculated from $N$-body simulations as
\begin{equation}
\bar{n}_\text{ray} = \left ( \frac{\Omega_\text{b}}{\Omega_\text{m}} \right ) 
\left ( \frac{\rho_\text{DM}}{\mu m_\text{p}} \right),
\end{equation}
where $\rho_\text{DM}$ is the mean dark matter density, $\mu = 1.2$ is the mean molecular weight of the neutral gas, and $m_\text{p}$ is the proton mass.  Our results suggest that most of the DC halos lie outside the HII region
 ($R > R_\text{st}$) for the source UV luminosity of $10^{41}~\mathrm{erg~s^{-1}}$. These halos remain shielded from ionizing photons  as they approach the source galaxy. The halos located close to $R=R_\text{st}$ may get affected from ionizing photons as shown by \cite{Regan+2016a,Regan+2016b}. Hence, we expect that ionizing photons do not significantly affect the collapse in most of the DC halos. 

In the  above analysis,  we have neglected hydrodynamical effects such as mechanical feedback from the source galaxy and locally compression of the gas by shocks or turbulence etc. How these processes affect the mean density between the DC halo and the source halo cannot be estimated from our semi-analytical model. In future, hydrodynamical simulations coupled with radiation transfer, like in section 3 of the current study,  are required to comprehend the role of these effects.

\subsection{Environment dependence of $R_\text{st}$} \label{sec_environment}
The size of  $R_\text{st}$ depends on the mean density of the environment around the source ($\bar{n}$, equation~\ref{eq_rst}) that is highly anisotropic  as shown in Fig.~\ref{fig_modelsfe01}. Here, we consider how  different environments affect the expansion of an ionized region.

In Fig.~\ref{fig_R_LUV}, we show the mean density of the halo and the density profiles in voids and filaments assuming that the central halo mass ($M_\text{halo}$) of $10^7~M_\odot$ at $z=27$. Here, the density within the virial radius is constant and equal to $\delta_\text{c} \bar{\rho}$ for all three cases, where $\delta_\text{c} \equiv 168$ and $\bar{\rho}$ is the cosmic mean density. The averaged density profile outside the virial radius is evaluated in two ways.  In the first method, we estimate the profile from the extended Press--Schechter and the spherical collapse models \citep{Barkana2004} while in the second method, we assume NFW profile \citep{NFW1997} outside the virial radius. In filaments, the density is assumed to be the same as the spherically averaged profile inside the radius of about 200 pc while outside the density is the fixed at  $(1 + \delta_\text{fil}) \bar{\rho}$. Here, we adopt $\delta_\text{fil} \equiv 100$ as found from our simulations.  In the void region, the density is set to be the cosmic mean density. The  comparison of density profiles show that  the density is  larger in filaments compared to the mean density of  the halo.

We also estimated the UV luminosity required to ionize neutral hydrogen within the radius $R$ from the source for various environments in the following way:
\begin{equation}
L_\text{UV} = \int_0^R 4\pi R^2 E_\text{UV}\alpha_\text{B} n^2 \mathrm{d}R .
\end{equation}
The critical UV luminosity for different environments is shown in Fig.\ref{fig_R_LUV}(b).  It is found that $L_\text{UV, crit}$ inside the viral radius is $L_\text{UV, crit} \sim 1\times 10^{40} ~\mathrm{erg~s^{-1}}$, remains roughly constant for voids and monotonically increases for filaments.  Our estimates suggest that $L_\text{UV, crit}$ in filaments is about  two orders of magnitude larger compared to the voids.

\begin{figure}
	\centering
	\includegraphics[width=7.3cm]{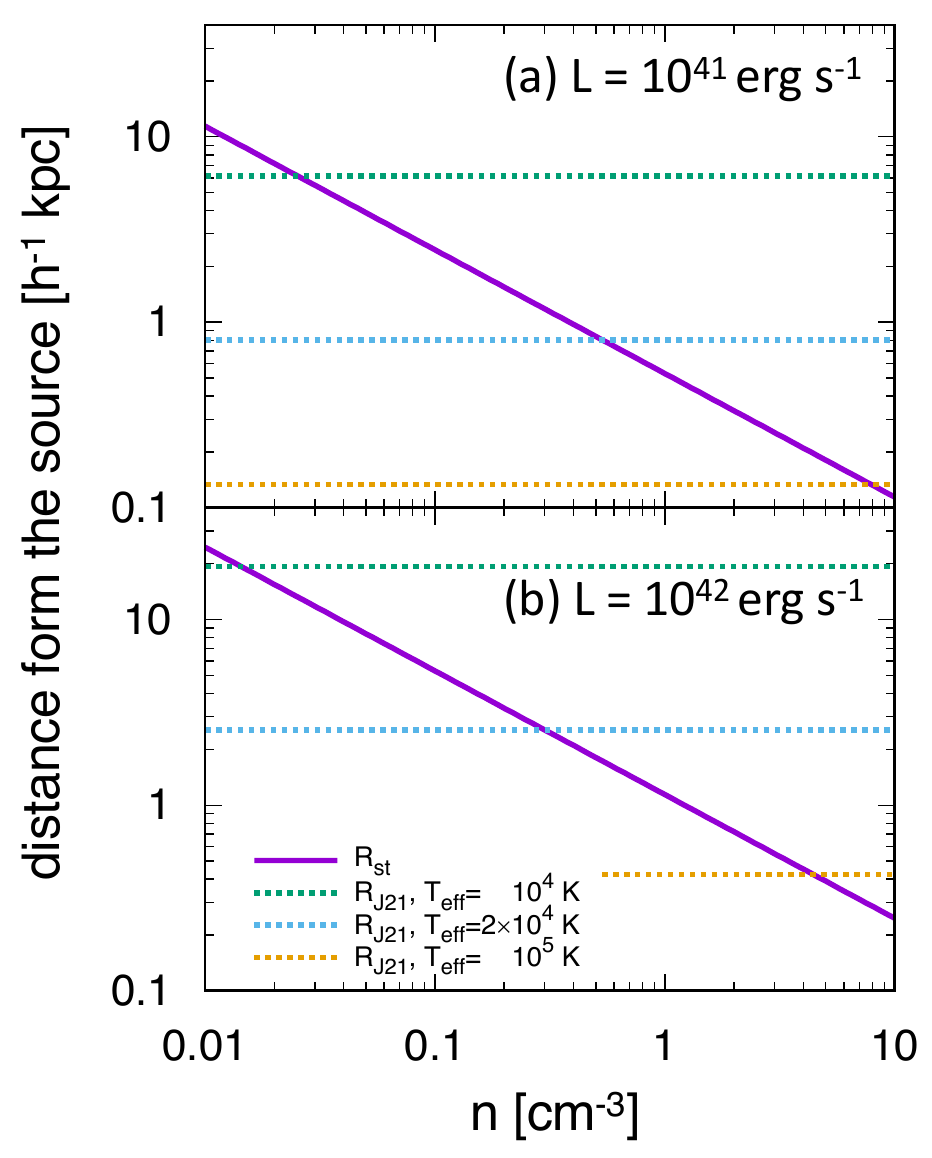}
	\caption{
	$R_\text{st}$ (equation \ref{eq_rst}) and $R_\text{J21}$ (equation \ref{eq_rJ21}) as a function of a mean number density ($\bar{n}$) under UV luminosity of (a) $10^{41}~\mathrm{erg~s^{-1}}$ and (b) $10^{42}~\mathrm{erg~s^{-1}}$, respectively. The purple lines show $R_\text{st}$, while dotted lines show $R_\text{J21}$ assuming 	the blackbody spectra with $T_\text{eff} = 10^4$ (green), $2\times10^4$ (blue), and $10^5~$K (yellow). $R_\text{J21}$ does not depend on $\bar{n}$ because we assume LW radiation is 	optically thin.
	}
	\label{fig_Rst_Rj21}
\end{figure}

\subsection{Spectrum of the radiation source} \label{sec::discussion_spectrum}
Recent studies suggest that realistic spectra of Pop II stars can be mimicked by a blackbody radiation spectrum between $\rm 10^4$ and $10^5$ K \citep{Agarwal+2014, Sugimura+2014,  Latif+2015,WolcottGreen+2016}. They found that $J_{21,\text{crit}}$ converges to $\sim 1000$ for spectra harder/equal to T2e4, while for T1e4 spectrum $J_{21,\text{crit}} = 30$--$100$.  In the following, we discuss how the choice of the radiation spectrum affects the size of HII region and  our conclusions about the impact of ionization on the collapse of DC halos.

\begin{table}
	\centering
	\scalebox{1.2}[1.2]{
	\begin{tabular}{ccccc} \\[2pt] \hline \hline  
		                      & T1e4 & T2e4 & T1e5 \\[2pt] \hline \\[0.25pt] 
	$J_{21,\text{crit}}$ & 100   &  1000 & 1000 \\[2pt]
	$\beta$                  & 8.43  & 1.45   & 0.04  \\[2pt] \hline \hline
	\end{tabular}
	}
	\caption{Summary of $J_{21,\text{crit}}$ and $\beta$.}
	\label{tab_spectra}
\end{table}

We consider three types of the blackbody spectra with $T_\text{eff} = 10^4$ (T1e4), $2\times10^4$ (T2e4), and $10^5~\mathrm{K}$ (T1e5). These spectra represent the old galaxy, the starburst galaxy, and the Pop III cluster, respectively.
The estimates of  $\beta$ (the ratio between LW and ionizing luminosities) for various radiation spectra along with $J_{21,\text{crit}}$ are listed in Table~\ref{tab_spectra}.

We show the evolution of $R_\text{st}$ and $R_\text{J21}$ against the mean number density ($\bar{n}$) in Fig.~\ref{fig_Rst_Rj21}  for  T1e4, T2e4, and T1e5 spectra, respectively. $R_\text{J21}$ does not depend on $\bar{n}$ because we assume LW radiation is optically thin (see Section~\ref{sec_stromgren}). There is a critical density $\bar{n}_\text{c}$, at which $R_\text{st}$ becomes equal to $R_\text{J21}$ and this value gets larger as spectrum becomes harder.  For T2e4 spectrum, $\bar{n}_\text{c} \sim 0.3$ -- $0.5~\mathrm{cm}^{-3}$, which corresponds to the filament density (Fig.~\ref{fig_modelsfe01}) while for T1e5 spectrum, $\bar{n}_\text{c} \sim 10~\mathrm{cm}^{-3}$,  corresponds to the halo density in our simulation. This suggests that DC halos located in the void region are likely captured by the HII region while the halos forming in a clustered environment (surround by dense filaments) may survive the impact of ionizing radiation for T2e4 spectrum. However, for T1e5 spectrum, DC halos are captured by the HII region even though they approach through the filament with $\sim 1~\mathrm{cm^{-3}}$.  Hence,  the formation of  an SMS  in the close vicinity of Pop III might be difficult compared to the Pop II stars.

We assumed $10^4~$K blackbody spectrum for LW radiation and $10^5~$K for ionizing photons. The adopted spectrum for LW is softer than the expected spectra of Pop II galaxies and the required stellar mass would be about an order of magnitude larger \citep{Sugimura+2014}.  The strength of critical LW flux required to enable isothermal DC collapse is 1000 for a realistic spectrum of Pop II galaxies, about an order of magnitude larger  than the adopted value. This will significantly reduce the number of candidate DC halos.  However,  the radiation spectrum used in our simulation has $\beta \sim 1$ and  the LW intensity  reaches $J_{21} \sim10^3$ during the collapse of DC halo. Therefore, our results are similar to $2\times10^4~$K blackbody spectrum.

\section{Conclusions}
We have performed SPH simulations coupled with radiative transfer to investigate the effect of ionizing radiation on the collapse of DC halos.  The simulated halos are selected from  \cite{Chon+2016} and UV luminosity is computed from the  semi-analytical model with different star formation efficiencies, $\alpha = 0.005$ and $0.1$, where latter is the maximum value compatible with the observation of the luminosity function at $z=6$--$7$. In both cases, the cloud within the DC halo remains neutral and collapses into a dense core. We found that the dense filaments  formed around the source galaxy  protect the DC halo from ionizing radiation. Interestingly, for the case of $\alpha=0.1$, the cloud collapses more rapidly than for the case with $\alpha=0.005$. This suggests that cloud collapse gets accelerated due to the photoheating of surrounding gas.

To compare our findings with the previous study \citep{Regan+2016a}, we  investigated the evolution of the DC halo for constant luminosities of $L_\text{UV} = 10^{41}~\mathrm{erg~s^{-1}}$, the same as in their work, and $10^{42}~\mathrm{erg~s^{-1}}$.
We found that for the case of $L_\text{UV} = 10^{42}~\mathrm{erg~s^{-1}}$, the SMS-forming cloud is photoheated up to a $\text{few } \times 10^4~$K and gets completely evacuated. For the case of $L_\text{UV} = 10^{41}~\mathrm{erg~s^{-1}}$, one of the DC halo was continuously heated by ionizing radiation, could not collapse  and finally merged with the source galaxy. While another DC halo was able to collapse as it wmed inside the dense filament.  Our results suggest that  some of the DC halos can collapse even under strong ionizing radiation ($L_\text{UV} = 10^{41}~\mathrm{erg~s^{-1}}$) at the early stage of halo formation.
The luminosity at the end of simulation ($t\sim0.1~$Gyr) of the case with $\alpha=0.005$ is comparable to $L_\text{UV} = 10^{41}~\mathrm{erg~s^{-1}}$. Thus the difference between the cases with $\alpha=0.005$ and constant luminosity with $L_\text{UV} = 10^{41}~\mathrm{erg~s^{-1}}$ demonstrates that the luminosity evolution of the source galaxy is important.

We also estimated  the effect of ionizing radiation on 68 DC halos found in \cite{Chon+2016} by employing the semi-analytical model. Our results suggest that as a DC halo approaches the radiation source, the mean density between the source galaxy and the DC halo increases that protects it from the ionizing radiation. The estimated size of the HII region remains smaller, at least for one dynamical time, in most of the cases. This picture is consistent with our specific sample where we performed radiation hydrodynamical simulations.  Thus, we conclude that in most cases ionizing radiation is shielded by the filamentary structure around the source galaxy and does not prevent the formation of  an SMS.

\section *{Acknowledgement}
We thank K. Sugimura, G. Chiaki, Y. Sakurai, T. Okabe, R. Nakatani, T. Hosokawa, N. Yoshida, and J. Regan for fruitful discussions and comments. This work was financially supported by Advanced Leading Graduate Course for Photon Science program me and by the Grants-in-Aid for Scientific Research from the JSPS Promotion of Science (16J07507). This project has received funding from the European Union’s Horizon 2020 research and innovation programme under the Marie Sklodowska-Curie grant agreement No 656428. The numerical simulations were carried out on XC30  at the Center for Computational Astrophysics (CfCA) of  National Astronomical Observatory of Japan. We used the SPH visualization tool SPLASH \citep{SPLASH} in Figs~\ref{fig_modelsfe01} and \ref{fig_model1e41}.

\bibliography{biblio3}

\bsp	
\label{lastpage}
\end{document}